\documentclass[conference]{IEEEtran}
\IEEEoverridecommandlockouts
\usepackage{cite}
\usepackage{amsmath,amssymb,amsfonts}
\usepackage{algorithmic}
\usepackage{graphicx}
\usepackage{subcaption}
\usepackage{textcomp}
\usepackage{xcolor}
\usepackage[hyperfootnotes=false]{hyperref}
\usepackage{verbatim}
\usepackage{dblfloatfix}
\usepackage{enumitem}
\def\BibTeX{{\rm B\kern-.05em{\sc i\kern-.025em b}\kern-.08em
    T\kern-.1667em\lower.7ex\hbox{E}\kern-.125emX}}

\newcommand\blfootnote[1]{%
  \begingroup
  \renewcommand\thefootnote{}\footnote{#1}%
  \addtocounter{footnote}{-1}%
  \endgroup
}

\begin{document}

\title{Why we care \\
(about quantum machine learning)}

\author{
  \IEEEauthorblockN{
    Richard A. Wolf\\[2mm]
  }
  \IEEEauthorblockA{
    Irish Centre for High-End Computing (ICHEC)\\
    University of Galway
  }
}

\maketitle

\blfootnote{For correspondence, write to \href{mailto:richard.a.wolf@proton.me}{richard.a.wolf@proton.me}.}

\begin{abstract}
    Quantum machine learning has received tremendous amounts of attention in the last ten years, and this trend is on the rise. Despite its developments being currently limited to either theoretical statements and formal proofs or small-scale noisy experiments and classical simulations, this field of quantum technologies has been consistently standing in the spotlight. Moreover, the locus of attention seems to have been skewed towards three central questions: "Can we beat classical computers?", "How?" and "When?". In this work, I argue that focus on quantum machine learning stems from a wide range of factors, some of which lie outside the discipline itself. Based on both recent and key publications on the subject as well as general audience sources, I give a brief overview of the core questions being raised in quantum machine learning and propose a socio-epistemologic interpretation of the motivations behind those and interplay between them.
\end{abstract}


\section{Introduction}

Quantum machine learning (QML) has been one of the subfields of quantum sciences and technologies (QSTs) receiving the most spotlight in recent years. Beyond interest from specialists and colleagues from neighbouring disciplines, QML has also received sustained consideration from industry and governments.

This extended abstract intends to raise general awareness in the quantum computing (QC) and more specifically in the QML community around the possible reasons for the field's success lying outside its scientific and technical aspects. The aim of this work is to raise a series of questions, about QML and QSTs at large from an epistemological and socio-anthropological perspective. It by no means aspires to provide a definite \textit{answer} to any of those questions and only suggests food for thought which I hope will be taken further by the readers.

\section{Methods}

In order to isolate questions of interest and examine them through an epistemologic and socio-anthropological perspective, the following steps were taken:

\begin{itemize}
    \item Analyse a corpus of 10 specialised publications (both key by citation count, and recent) in the field of QML including abstract, introductions and conclusions.
    \item Analyse a corpus of 5 headlines and synopsis of quantum-machine learning related general-audience articles taken from recent sources found with keywords \textit{quantum machine learning (\_\_ / industry / policies/}).
    \item Analyse those through the lens of existing work on epistemology as well as philosophy and sociology of science.
\end{itemize}

The details of the above-mentioned corpora can be found in annex \ref{sec:annex}.

In order to tackle these questions, I first give a brief overview of the difference between \textit{beliefs} and \textit{knowledge} based on the definitions adopted in the previous section.

\section{Analysis and discussion}
\subsection{The fine line between \textit{knowledge} and \textit{belief}}

     An in-depth analysis of whether stances about QML held are within the realm of knowledge or belief belongs to epistemology and lies outside the scope of this work. Here, I will limit myself to particular pre-existing definitions of those concepts to classify the attitude towards QML.

    In this work I will resort to one of the definitions of \textit{belief} proposed by sociology and anthropology as the attitude of regarding some proposition as true. As for \textit{knowledge}, since the minimalist \textit{justified true belief} (JTB) definition has been widely accepted to be insufficient \cite{gettier1963justified}, I shall consider as a basis its \textit{no-defeater} modification whereby knowledge is defined as undefeated JTB \cite{lehrer1969knowledge}. Though this definition of knowledge as undefeated JTB is still considered insufficient in several instances \cite{olen1976undefeated} and a wide array of newer revised interpretations have been proposed, I shall restrict myself to its use for the purpose of this brief work based on the relative ease of analysis of each of its components. I however invite the readers to explore further definitions of knowledge and confront them to their understanding of the different fields of QTs \cite{al2020linking, goldman2021reliabilist} as undefeated JTB only allows us to outline general coarse-grained aspects of what defines knowledge but by no means suffices.

    Unpacking this definition of knowledge leads us to consider whether the belief in QML is:
    \begin{itemize}
        \item Justified.
        \item True.
        \item Undefeated.
    \end{itemize}

    The need for \textit{justification} arises for any piece of non-basic knowledge.Basic knowledge can be known to be true in and by itself, without relying on any other statement \cite{lehrer1969knowledge} while non-basic knowledge is derived from the truth value of other statements. In this context, a true belief can be defeated by one of its building-bloc statements being defeated. Given the complexity of the QML field, beliefs are here non-basic and based on the composition of sub-statements which are assumed to be true. In this context, the question of undefeatedness can be derived from the JTB analysis of each of its composing parts.

    Now, analysing the truth and justification aspects requires to take stances on a number of points. One can wonder about the validity of experiments involving simulations alone \cite{barberousse2009computer} or the legitimacy of quantum hardware-based experiments at scales and noise levels which are insufficient. It is also possible to question the accuracy of scaling projections -based on those small-scale experiments- and the soundness of formal theoretical statements \cite{frans2014mathematical}. These steps are crucial in ensuring falsifiability of QST, a necessary condition for valid science \cite{popper1963science}.

    Though answers to the questions raised in this section would no doubt prove satisfying to read, the brevity of this work would only afford hand-wavy and incomplete -if not outright innacurate- responses. Instead of providing coarse-grained responses, this section simply aspires at providing a minimal toolbox for the reader to analyse QML beliefs through their own lenses and based on their own interpretation of each of these points. Moreover, should the reader desire to interrogate these beyond the raw lens of undefeated JTB, the possibility to explore definitions of knowledge in the field of epistemology should prove to be a satisfying goal.
\subsection{Questions raised in QML}
Though the discipline enjoys a vast array of questions in the thousands of papers published in the field, some core interrogations seem to recur more often than others. Among these, one finds:
  \begin{itemize}
      \item \textbf{Timeline} of quantum computers capable of running quantum machine learning (QML) algorithms.
      \item \textbf{Applications} most likely to benefit from QML.
      \item \textbf{Performance} gains which can be reached.
      \item \textbf{Complexity} advantages over classical computations \cite{huang2021information, cerezo2021variational}? Limits of those \cite{aaronson2015read, tang2021quantum, schuld2022quantum}
      \item \textbf{Accuracy} \cite{huang2021information, cerezo2021variational}
      \item \textbf{Trainability} \cite{cerezo2021variational} and \textbf{generalisation} \cite{peters2022generalization} 
  \end{itemize}
  
\subsection{The rationality behind QML's support}

    More often than not, science boasts of holding epistemologically rational beliefs as opposed to pragmatically rational ones. \textit{Epistemological rationality}, sometimes also termed \textit{evidential rationality}, refers to how well beliefs map onto the real world and the experience we have of it \cite{buchak2010instrumental, eder2021explicating}. On the other hand, \textit{pragmatic rationality} refers to how well beliefs align with the pursuance of certain goals \cite{buchak2010instrumental}. I argue that considering the possibility that belief in QML could be pragmatic rather than purely epistemological could shift the focus of attention from the \textit{causes} of this belief to their \textit{consequences}. By de-coupling the soundness of holding such belief with the consequences holding it has, it might be possible to make a case in favour of, or against, such belief, independently of its truth value.

    \subsubsection{Positive consequences}
    A number of positive consequences can be causally linked to the belief -whether pragmatic or not- in QML. Through the spotlight it is under, it benefits from increased job opportunities and interest both from the QST community and general audience. This in turn, increases possibilities for funding and investment. Those aspects could potentially lead to a form of self-fulfilling prophecy \cite{merton1948self} where QML's potential for success might end-up deriving from the attention it was given instead its success being the reason for said belief.

    \subsubsection{Negative consequences}
    However, not all consequences of the enthusiasm for QML are positive and any discipline receiving increased attention can also be subject to more strain or cause rippling effects to its neighbours.
    A first risk stems from the temptation of turning a blind eye on potential negative or at least mitigated results, hyper-tailoring problems to force positive results, irrelevant of their usefulness. Another risk is the temptation to \textit{quantise} parts of algorithms for the sake of quantising them without any real improvement over classical counterparts. I argue that the use of quantum computing should, at least for now, be reserved to applications which \textit{need} \textit{quantumness} in some aspect. By this, I mean avoiding to use quantum because we \textit{can} and only using it because we \textit{need} to. When choosing between a classical and a quantum algorithm, the motivation for quantum should be based on the impossibility of performing a certain task classically with given benefits, rather than on the possibility to perform it quantumly. Those benefits do not need to be purely in terms of computational complexity but could be based on other factors such as for instance energy consumption \cite{meier2023energy}. Another potential negative consequence can be seen when looking beyond QML itself and focusing on the larger field of QSTs.
    The second likely negative consequence of the belief in QML stems from the self-fulfilling prophecy mentioned above can. The simple reason behind this is the finiteness of financial and human resources available to QSTs as a pool: the resources that go into QML are \textit{not} going into other sub-fields. While this is not necessarily an issue, it should nevertheless be taken into account when analysing the popularity and performance of the sub-field. One should bear in mind the advances that could potentially be made in other sub-fields if they benefited from the same human and financial attention. If QML is to triumph in any way, it should do so in a \textit{fair} competition with other sub-fields and not simply by diverting all resources towards itself. While it is of course clear that a portion of the human and financial efforts currently being invested in QML would still be targeted at it were its popularity lesser, it is also true that a possibly significant part might go to other sub-fields of QST if they were more prominent in the general minds. A graduate student looking for an internship in QST might end up in QML because of the increased number of opportunities available, while they might have benefited from being exposed to foundations of quantum information instead. A company looking to start investing in QSTs might be impressed by the visibility of QML and decide to focus there when they could have potentially leaned towards complexity theory.

    \subsubsection{Confusion between pragmatic and epistemic rationality}
    Beyond the potential pitfalls, the problem of holding a potentially pragmatically rational belief in QML comes from posing it as epistemological. Currently, few members of the QML community acknowledge the possibility that their rationality could be functional rather than epistemologic, creating confusion between causes and consequences of such belief. While there is nothing inherently wrong with either kind of rationality, I suggest that being clear on its actual type could lead to a better understanding of the sub-field and its relation to other subfields of QST.
\subsection{QML as a token of the modern sacred}

    QML can easily be thought of as the quantum cousin of classical machine learning (ML). Thanks to this, people were able to build imaginary representation of QML based on the pre-existing ones for ML. This was particularly useful for the general audience who might have little to no access to scientific works and could therefore focus on the popular culture elements around classical ML -often referred to as \textit{Artificial Intelligence} (AI) in its popular forms-. Because few science fields have been as explored as AI in works of fiction and popular culture \cite{arnold2017turing, goode2018life}, other sub-fields of QST could not claim the same wide popularity of pre-existing cousins. Through this, QML benefited from pre-existing collective imaginary built on ML and artificial intelligence -terms often used interchangeably in general media-, making it more salient.
      
    I argue that from a sociological and anthropological perspective, the belief in QML's possibilities is a token instance of the more general relationship to science and deep technologies as modern, more acceptable \cite{foucault2007security}, instances of the sacred. 
    In this work I resort to Durkheim's definition of the sacred which is considered as being both \textit{set apart} from the profane and \textit{forbidden} or inaccessible \cite{durkheim2016elementary}. To this, I add the presence of a sacerdotal dedicated class endowed with the power to interface with the sacred, present in anthropology.
    
    QSTs at large seem to fulfill the condition of inaccessibility, with not only general audience but also researchers from other STEM fields accepting their impossibility to interact with the \textit{quantumness}. Even more, experts of the field itself have cultivated the myth of inaccessibility, with quotes like Feynmann's \textit{if you think you understand quantum mechanics, you don't understand quantum mechanics} or Einstein's \textit{spooky action at a distance}. Whether those quotes are accurately attributed or not is irrelevant, they are often assumed to be so, setting quantumness as inaccessible, both for the general general audience as well as for the experts. Separateness from the profane -or classical- is here another key similarity.
    
    Quantumness encompasses everything that is classical \textit{and more}, emphasising its poietic power. This idea could potentially be linked to animist views of a \textit{something} that inhabits all and transcends it, which could in this case be assimilated to quantumness.
    
    Finally, \textit{experts} can be equated to a dedicated sacerdotal class, capable to some extent and despite its inaccessibility to interact with quantumness as the sacred object.

\section{Conclusions, limitations and future work}

\subsection{Conclusions, limitations}
This extended abstract only scratched the surface of questions which deserve more sustained and in-depth analysis. Nevertheless, I hope it will have served in sparking an interest in those questions, prompting the reader to follow-up in exploring some of those interrogations.
One of the key conclusions is that some part of the rationality behind the support of QML appears to be pragmatic, despite often being interpreted as epistemic. Another key idea surfaces when looking at QML as a token of general \textit{deep techs}. In this context, it seems possible to read it as an element of the modern sacred.

\subsection{Future work}
A compilation of short-listed questions worthy of interest is given below. A proposed approach to tackle these would be to go beyond corpus-based analysis and leverage both surveys and semi-structured interviews with core actors of the field. Another possible road would be to explore rigorous argument construction and question it in light of philosophy of knowledge and philosophy of science.

\begin{enumerate}
    \item To which extent does the focus on \textit{possibilities}, as built by the imaginary around CML be a contributing factor to QML's popularity? 
    \item What is the interplay between pragmatic rationality and economics in the context of QML? How can this be analysed through a technology-investment game-theoretic approach \cite{huisman2013technology, zhu2003strategic, abdou2018battle}?
    \item How can the support for QML be understood in the context of a rhetoric of innovation \cite{joly2015regime}? And in this context, what are its true efforts towards embedding itself in a meaningful social context\cite{joly2015regime}?
    \item How does the use of emotion-based vocabulary in abstracts and introductions of QML papers contribute to its popularity from a discourse analysis perspective? Which specific affects are being leveraged through such lexicon? What kind of bond people created with QML and how could this be understood as the new type of social bonding beyond the human-human and towards the human-technological object \cite{latour2020aramis}?
    \item How does the myth of inaccessibility of quantum relate to the modern sacred and how does it contribute to demographics selection in the field?
\end{enumerate}

\section*{Acknowledgments}
Warm thanks you to my amazing reviewers who supported this project by providing they precious feedback. Your help was inestimable dear Sara Metwalli and everyone else.

\bibliographystyle{IEEEtran}  
\bibliography{main}  

\section*{Annex}\label{sec:annex}

\subsection*{Specialist sources}
\footnotesize{
\begin{enumerate}
    \item \textit{Quantum Machine Learning in Feature Hilbert Spaces} \cite{Schuld2018QuantumML}.
    \item \textit{Quantum algorithm for linear systems of equations} \cite{harrow2009quantum}.
    \item \textit{An introduction to quantum machine learning} \cite{schuld2015introduction}.
    \item \textit{Generalization despite over-fitting in quantum machine learning models} \cite{peters2022generalization}.
    \item \textit{Challenges and opportunities in quantum machine learning} \cite{cerezo2022challenges}.
    \item \textit{Information-theoretic bounds on quantum advantage in machine learning} \cite{huang2021information}.
    \item \textit{Variational quantum algorithms} \cite{cerezo2021variational}.
    \item \textit{Read the fine print} \cite{aaronson2015read}.
    \item \textit{Quantum principal component analysis only achieves an exponential speedup because of its state preparation assumptions} \cite{tang2021quantum}.
    \item \textit{Is quantum advantage the right goal for quantum machine learning?} \cite{schuld2022quantum}
\end{enumerate}
}

\subsection*{General audience sources}
\footnotesize{
\begin{enumerate}
    \item \textit{Artificial Intelligence, Quantum Computing, and Space are 3 Tech areas to Watch in 2024} \cite{website1}
    \item \textit{2024 will see convergence of GenAI, quantum computing} \cite{website2}
    \item \textit{When AI meets Quantum Computing} \cite{website3}
    \item \textit{Will EU leaders take up the quantum challenge in front of them?} \cite{website4}
    \item \textit{Quantum Artificial Intelligence Is Closer Than You Think} \cite{website5}
\end{enumerate}
}

\end{document}